\documentclass[pre,showpacs]{revtex4}
\usepackage{graphicx}

\begin{document}

\title{A fractional diffusion equation for two-point probability distributions of a continuous-time random walk}

\author{A. Baule$^1$ and R. Friedrich$^2$}

\affiliation{ $^1$School of Physics and Astronomy,
University of Leeds, 
LS2 9JT,
United Kingdom\\
$^2$Institute for Theoretical Physics, University of M\"unster, Wilhelm-Klemm Str. 9, 48149 M\"unster, Germany
}

\begin{abstract}

Continuous time random walks are non-Markovian stochastic processes, which are only partly characterized by single-time probability distributions. We derive a closed evolution equation for joint two-point probability density functions of a subdiffusive continuous time random walk, which can be considered as a generalization of the known single-time fractional diffusion equation to two-time probability distributions. The solution of this generalized diffusion equation is given as an integral transformation of the probability distribution of an ordinary diffusion process, where the integral kernel is generated by an inverse L\'evy stable process. Explicit expressions for the two time moments of a diffusion process are given, which could be readily compared with the ones determined from experiments.

\end{abstract}


\maketitle

The concept of continuous time random walks (CTRWs), introduced by Montroll and Weiss \cite{Montroll} almost four decades ago, has been successfully applied to a wide variety of transport problems in physics \cite{Bouchaud}. In recent years the relationship between CTRWs and a class of Fokker-Planck equations with fractional temporal and/or spatial derivatives has attracted a lot of attention \cite{Metzler1}. In this line of research a major focus has been on the investigation of anomalous diffusion \cite{Shlesinger}, a nonequilibrium phenomenon occurring in scientific fields ranging from astrophysics to biophysics and econophysics. A well known example is a subdiffusive CTRW in a force field, which leads to the well-known \textit{fractional Fokker-Planck equation} (FFPE) \cite{Metzler2}, invoking the fractional Riemann-Liouville differential operator \cite{Podlubny}. This equation can be written in the equivalent form:
\begin{eqnarray}
\label{Intro_FFPE}
_0C_t^\alpha f(x,t)=L_{FP}f(x,t),
\end{eqnarray}
where $_0C_t^\alpha$ denotes the Caputo fractional differential operator \cite{Podlubny}, defined as
\begin{eqnarray}
\label{Intro_Caputo}
_0C_t^\alpha g(t)\equiv\frac{1}{\Gamma(1-\alpha)}\int_0^t g^{(1)}(t')(t-t')^{-\alpha}
\end{eqnarray}
for $0<\alpha<1$ ($g^{(1)}(t)$ is the first derivative of g(t)). $L_{FP}$ is a Fokker-Planck operator. For simple diffusion processes in $1d$ $L_{FP}=D\partial^2/\partial x^2$ and Eq.~(\ref{Intro_FFPE}) is referred to as \textit{fractional diffusion equation} (FDE) \cite{Schneider, Podlubny}. In the limit $\alpha\rightarrow1$ Eq.~(\ref{Intro_FFPE}) reduces to the ordinary Fokker-Planck equation. Clearly, the Caputo fractional time derivative expresses the non-Markovian character of the underlying CTRW: the probability of finding the random walker at point $x$ at time $t$ depends on the whole history of the process from time $0$ up to $t$. In combination with fractional evolution equations, the CTRW constitutes a versatile stochastic model which can take into account sub- and superdiffusive behaviour in complex systems as diverse as turbulence \cite{Friedrich1}, optical lattices, and biological cell motility (see \cite{Metzler3} for an extensive review). Recent new developments concern a more fundamental understanding of CTRWs and their application, e.g. the connection to ageing phenomena \cite{Barkai}, ergodicity breaking \cite{Bel} and inertial particles diffusing in a potential \cite{Friedrich2}.

However one important aspect has long been neglected in the literature on CTRWs and FFPEs. In general the ordinary Fokker-Planck equation is a deterministic evolution equation for the \textit{transition probability} of the stochastic process, which is assumed to be Markovian. By virtue of the Markovian property one can calculate arbitrary $n$-point pdfs with a single transition probability. On the other hand, in the case of the fractional analog Eq.~(\ref{Intro_FFPE}), \textit{single-time} pdfs are determined, which contain only very limited information about the stochastic process. In order to obtain complete information about a non-Markovian process like the CTRW, one has to consider the infinite set of $n$-point pdfs. Recently, such multi-point statistics have been investigated in different approaches \cite{Barsegov,Baule}. In \cite{Barsegov}, a two-point Green function has been determined with the help of the FFPE backward propagator and allows for the computation of three-point fluorescence lifetime correlation functions. Our previous work \cite{Baule} discussed joint probability distributions of a CTRW based on a representation in terms of coupled Langevin equations. The purpose of the present paper is to derive a closed evolution equation for the two-point pdf of a CTRW, which is independent of the inverse L\'evy-stable process investigated in \cite{Baule}. This equation is the generalization of the single-time FDE (\ref{Intro_FFPE}) to two times. Considering the widespread interest in Eq.~(\ref{Intro_FFPE}), the two-point generalization is expected to provide important further insight into the application of CTRWs to real world systems. In fact, multi-point statistics have already been probed in recent experiments on protein conformational dynamics \cite{Yang,Min} and highlight the importance of a more complete theoretical understanding of these statistics in anomalous diffusive systems. 

A suitable starting point for the investigation of CTRWs is a representation in terms of coupled Langevin equations introduced by Fogedby \cite{Fogedby}. Here the motion of a Brownian particle in an external force field $F(x)$ in $d=1$ dimensions is described as:
\begin{eqnarray} \frac{dX(s)}{ds}&=& F(X) + \eta(s),
\label{Langevin1}\\ \frac{dt(s)}{ds}&=& \tau(s). \label{Langevin2}
\end{eqnarray}
In this framework the CTRW is parametrized by the continuous path variable $s$, which may be regarded as arclength along the trajectory. $X$ denotes the physical space and $t$ the physical time. Both are given as stochastic processes in the 'eigentime' $s$. Their statistics are determined by the properties of the stochastic variables $\eta(s)$ and $\tau(s)$. In this work we consider the special case of statistically independent increments $\eta$ and $\tau$, i.e. jump lengths and waiting times are uncoupled. Furthermore we want to restrict our considerations to a \textit{subdiffusive} CTRW without force field. Accordingly, $\eta(s)$ is assumed as a standard Langevin force with properties $\left<\eta(s)\right>=0$ and $\left<\eta(s)\eta(s')\right>=\zeta\delta(s-s')$ . $X(s)$ is consequently given as a standard Markovian Wiener process. The subdiffusive characteristics enter via the process $t(s)$. Its increments $\tau(s)$ are assumed to be broadly distributed such that $t(s)$ constitutes an asymmetric L\'evy-stable process of order $\alpha$ with $0<\alpha<1$. L\'evy-stable processes of this kind induce a diverging characteristic waiting time $\left<t(s)\right>$. We are interested in the process $X(t)=X(s(t))$, i.e. the behaviour of the physical space variable $X$ as a function of physical time $t$. The pdfs of this process are defined as (e.g. for two points):
\begin{eqnarray}
\label{2_pdf_def}
f(x_2,t_2;x_1,t_1)=\left<\delta(x_2-X(s_2))\delta(s_2-s(t_2))\delta(x_1-X(s_1))\delta(s_1-s(t_1))\right>.
\end{eqnarray}
As a consequence of above specifications, $X(t)$ is non-Markovian and reveals subdiffusive characteristics: $\left<x(t)^2\right>\sim t^\alpha$. Its statistical properties are closely related to the properties of the process $s(t)$, the inverse of the process $t(s)$. The investigation of the statistics of the inverse L\'evy-stable process $s(t)$ has been the focus of reference \cite{Baule}. One of the main results is the following integral transformation for the $n$-point pdf $f(\{x_i,t_i\})$ of the process $X(t)$ \cite{Baule}:
\begin{eqnarray}
\label{path_integral}
f(\{x_i,t_i\})&=&\int_0^\infty ds_1...\int_0^\infty ds_n\;h(\{s_i,t_i\})f_M(\{x_i,s_i\}).
\end{eqnarray}
$h(\{s_i,t_i\})$ denotes the $n$-point pdf of the process $s(t)$ and $f_M(\{x_i,s_i\})$ the $n$-point pdf of the Markovian process $X(s)$. Eq.~(\ref{path_integral}) states that the pdf of the non-Markovian process can be determined by a transformation of the corresponding Markovian process. The integral kernel is generated by the inverse L\'evy-stable process $s(t)$. Whereas the pdf $f_M(\{x_i,s_i\})$ is obtained in a straightforward way from the Langevin equation (\ref{Langevin1}), the determination of $h(\{s_i,t_i\})$ proves to be the crucial point. In the one- and two-point case $h$ assumes a simple form in Laplace space.

In the following we present the derivation of the two-point fractional diffusion equation starting from the transformation Eq.~(\ref{path_integral}). The derivation of the single-time FDE (\ref{Intro_FFPE}) proceeds along the same lines and is not shown here.
We define Laplace transforms in the usual way: $\mathcal{L}\{g(t)\}=\int_0^\infty g(t)e^{-\lambda t}dt$, accordingly for multiple variables $t_i\rightarrow\lambda_i$. Functions with argument $\lambda_i$ always denote the Laplace transform if not otherwise indicated: $g(\lambda)\equiv\mathcal{L}\{g(t)\}$. In Laplace space Eq.~(\ref{path_integral}) for the two-point pdf $f(x_2,t_2;x_1,t_1)$ assumes the similar expression:
\begin{eqnarray}
\label{2_solution_L}
f(x_2,\lambda_2;x_1,\lambda_1)&=&\int_0^\infty ds_1\int_0^\infty ds_2\;h(s_2,\lambda_2;s_1,\lambda_1)f_M(x_2,s_2;x_1,s_1).
\end{eqnarray}
The Laplace transform of the two-point pdf $h$ is explicitly given as \cite{Baule}:
\begin{eqnarray}
\label{2_h_L}
h(s_2,\lambda_2;s_1,\lambda_1)&=&\delta(s_2-s_1)\frac{\lambda_1^\alpha-(
\lambda_1+\lambda_2)^\alpha +\lambda_2^\alpha}{\lambda_1
\lambda_2}e^{-s_1(\lambda_1+\lambda_2)^\alpha} \nonumber \\
&&+\Theta(s_2-s_1)\frac{(\lambda_2^\alpha)
((\lambda_1+\lambda_2)^\alpha-\lambda_2^\alpha)} {\lambda_1 \lambda_2}
e^{-(\lambda_1+\lambda_2)^\alpha s_1}e^{-\lambda_2^\alpha (s_2-s_1)}\nonumber \\
&&+\Theta(s_1-s_2)\frac{(\lambda_1^\alpha)
((\lambda_1+\lambda_2)^\alpha-\lambda_1^\alpha)}
{\lambda_1 \lambda_2}
e^{-(\lambda_1+\lambda_2)^\alpha s_2}e^{-\lambda_1^\alpha (s_1-s_2)}.
\end{eqnarray}
An equation for $h(s_2,\lambda_2;s_1,\lambda_1)$ is readily derived in the form
\begin{eqnarray}
\label{2_heq_L}
&&\left(\frac{\partial}{\partial
s_1}+\frac{\partial}{\partial s_2}\right)
h(s_2,\lambda_2;s_1,\lambda_1)=-(\lambda_1+\lambda_2)^\alpha h(s_2,\lambda_2;s_1,\lambda_1),
\end{eqnarray}
with initial conditions
\begin{eqnarray}
\label{2_hinitial_L}
h(0,\lambda_2;0,\lambda_1)&=&\frac{\lambda_1^\alpha-
(\lambda_1+\lambda_2)^\alpha+\lambda_2^\alpha}
{\lambda_1 \lambda_2}, \nonumber \\
h(s_2,\lambda_2;0,\lambda_1)&=&\frac{(\lambda_2^\alpha)
((\lambda_1+\lambda_2)^\alpha-\lambda_2^\alpha)}
{\lambda_1 \lambda_2}e^{-\lambda_2^\alpha s_2}, \nonumber \\
h(0,\lambda_2;s_1,\lambda_1)&=&\frac{(\lambda_1^\alpha)
((\lambda_1+\lambda_2)^\alpha-\lambda_1^\alpha)}
{\lambda_1 \lambda_2}
e^{-\lambda_1^\alpha s_1}.
\end{eqnarray}
In general the two-point pdf $f_M(x_2,s_2;x_1,s_1)$ of the Markovian diffusion process is determined by the transition probability: $f_M(x_2,s_2;x_1,s_1)=P(x_2,s_2|x_1,s_1)f(x_1,s_1)$. In turn $P(x_2,s_2|x_1,s_1)$ is obtained from the ordinary Fokker-Planck equation. Due to the Markovian property it is not necessary to formulate a closed evolution equation for $f_M(x_2,s_2;x_1,s_1)$, however such an equation can be derived in a straightforward way from the two-point characteristic function of the process $X(s)$. The result is:
\begin{eqnarray}
\label{Joint_ODE}
\left(\frac{\partial}{\partial s_1}
+\frac{\partial}{\partial s_2}\right)f_M(x_2,s_2;x_1,s_1)= L(x_2,x_1)f_M(x_2,s_2;x_1,s_1),
\end{eqnarray}
where we define a generalized diffusion operator as the inverse Fourier transform
\begin{eqnarray}
\label{diff_op}
L(x_2,x_1)g(x_2,x_1)&\equiv&\mathcal{F}^{-1}\left\{-\frac{\zeta}{2}|k_1+k_2|^2 g(k_2,k_1)\right\}\\
&=&\frac{\zeta}{2}\left(\frac{\partial^2}{\partial x_1^2}+\frac{2\partial^2} {\partial
x_1\partial x_2}+\frac{\partial^2}{\partial x_2^2}\right)g(x_2,x_1).\nonumber
\end{eqnarray}
The definition Eq.~(\ref{diff_op}) holds in the same form in higher dimensions. Furthermore, broadly distributed symmetric increments $\eta(s)$ in Eq.~(\ref{Langevin1}) would result in a characteristic exponent $0<\mu<2$. The inverse Fourier transform then leads to a two-point generalization of the fractional Riesz/Weyl operator \cite{Metzler1}. The initial conditions of Eq.~(\ref{Joint_ODE}) read:
\begin{eqnarray}
\label{Joint_ODE_ic} f_M(x_2,s_2;x_1,s_1=0)&=&f_M(x_2,s_2)\delta(x_1), \nonumber \\
f_M(x_2,s_2=0;x_1,s_1)&=&f_M(x_1,s_1)\delta(x_2), \nonumber \\
f_M(x_2,s_2=0;x_1,s_1=0)&=&\delta(x_2)\delta(x_1).
\end{eqnarray}
Now we can proceed as follows. We multiply Eq.~(\ref{2_solution_L}) by $(\lambda_1+\lambda_2)^\alpha$ and substitute Eq.~(\ref{2_heq_L}). After performing partial integrations with respect to $s_1$ and $s_2$, Eq.~(\ref{Joint_ODE}) can be substituted leading to the following equation:
\begin{eqnarray}
\label{2_solution_bc1}
(\lambda_1+\lambda_2)^\alpha f(x_2,\lambda_2;x_1,\lambda_1)&=&L(x_2,x_1)f(x_2,\lambda_2;x_1,\lambda_1)+\int_0^\infty ds_2\:h(s_2,\lambda_2;0,\lambda_1)f_M(x_2,s_2)\delta(x_1)\\\nonumber&&+\int_0^\infty ds_1\:h(0,\lambda_2;s_1,\lambda_1) f_M(x_1,s_1)\delta(x_2).
\end{eqnarray}
The specific boundary terms occur due to the partial integrations and assuming $h(s_2 \rightarrow
\infty,t_2;s_1,t_1)=h(s_2,t_2;s_1 \rightarrow \infty,t_1)=0$ . These terms can be absorbed into the Caputo fractional differential operator, which we generalize to two times in a straightforward way. First we note that $h(s_2,\lambda_2;s_1=0,\lambda_1)$ and $h(s_2=0,\lambda_2;s_1,\lambda_1)$ are determined from Eq.~(\ref{2_h_L}):
\begin{eqnarray}
\label{2_hinitial_bc}
h(s_2,\lambda_2;0,\lambda_1)&=&\delta(s_2)\frac{\lambda_1^\alpha-(
\lambda_1+\lambda_2)^\alpha +\lambda_2^\alpha}{\lambda_1
\lambda_2}+((\lambda_1+\lambda_2)^\alpha-\lambda_2^\alpha)\frac{1}{\lambda_1} \lambda_2^{\alpha-1}e^{-\lambda_2^\alpha s_2}.
\end{eqnarray}
$h(s_2=0,\lambda_2;s_1,\lambda_1)$ takes the analogue form. Both expressions can be substituted into Eq.~(\ref{2_solution_bc1}). In the second term of Eq.~(\ref{2_hinitial_bc}) the Laplace transform of the single-time distribution $h(s_2,t_2)$ occurs \cite{Baule}:  $h(s_2,\lambda_2)=\lambda_2^{\alpha-1}e^{-\lambda_2^\alpha s_2}$. Consequently, with the help of Eq.~(\ref{path_integral}) for the case $n=1$ the integrations in the boundary terms of Eq.~(\ref{2_solution_bc1}) lead to the single-time distributions $f(x_2,\lambda_2)$ and $f(x_1,\lambda_1)$:
\begin{eqnarray}
(\lambda_1+\lambda_2)^\alpha f(x_2,\lambda_2;x_1,\lambda_1)
&=&L(x_2,x_1)f(x_2,\lambda_2;x_1,\lambda_1)+
((\lambda_1+\lambda_2)^\alpha-\lambda_2^\alpha)\frac{1}{\lambda_1}
f(x_2,\lambda_2)\delta(x_1)\nonumber\\&&+((\lambda_1+\lambda_2)^\alpha-\lambda_1^\alpha)\frac{1}{\lambda_2}f(x_1,\lambda_1)\delta(x_2)+2\frac{\lambda_1^\alpha-(
\lambda_1+\lambda_2)^\alpha +\lambda_2^\alpha}{\lambda_1
\lambda_2}\delta(x_2)\delta(x_1).
\end{eqnarray}
Rearranging terms yields:
\begin{eqnarray}
\label{2_ffpe_L}
&&\frac{(\lambda_1+\lambda_2)^\alpha}{\lambda_1\lambda_2}[\lambda_1\lambda_2 f(x_2,\lambda_2;x_1,\lambda_1)
-\lambda_2f(x_2,\lambda_2)\delta(x_1)-\lambda_1 f(x_1,\lambda_1)\delta(x_2)+\delta(x_2)\delta(x_1)]
\nonumber\\
&&\quad+ [\lambda_2^\alpha f(x_2,\lambda_2)-\lambda_2^{\alpha-1}\delta(x_2)]\frac{\delta(x_1)}{\lambda_1}+[\lambda_1^\alpha f(x_1,\lambda_1)-\lambda_1^{\alpha-1}\delta(x_1)]\frac{\delta(x_2)}{\lambda_2} \nonumber\\
&&=L(x_2,x_1)f(x_2,\lambda_2;x_1,\lambda_1)+\frac{\lambda_1^\alpha-(\lambda_1+\lambda_2)^\alpha+\lambda_2^\alpha}{\lambda_1
\lambda_2}\delta(x_2)\delta(x_1).
\end{eqnarray}
In order to perform the inverse Laplace transform of this equation we state the two results ($\Gamma(\beta)$ denotes the Gamma function):
\begin{eqnarray}
&&\mathcal{L}\left\{\Theta(t_2-t_1)\frac{t_1^{-\alpha}}{\Gamma(1-\alpha)}+\Theta(t_1-t_2)\frac{t_2^{-\alpha}}{\Gamma(1-\alpha)}\right\}=\frac{(\lambda_1+\lambda_2)^\alpha}{\lambda_1\lambda_2}, 
\end{eqnarray} 
and
\begin{eqnarray}
\mathcal{L}\left\{\frac{\partial}{\partial t_2}\frac{\partial}{\partial t_1}g(t_2,t_1)\right\}&=&\lambda_2\lambda_1 g(\lambda_2,\lambda_1)-\lambda_2 g(\lambda_2,t_1=0)-\lambda_1 g(t_2=0,\lambda_1)+g(t_2=0,t_1=0).
\end{eqnarray}
Furthermore, due to the definition Eq.~(\ref{2_pdf_def}) the boundary conditions of the pdf $f(x_2,t_2;x_1,t_1)$ are as in the Markovian case Eq.~(\ref{Joint_ODE_ic}): $f(x_2,t_2;x_1,t_1=0)=f(x_2,t_2)\delta(x_1)$, etc. The Laplace transform of the single-time Caputo operator Eq.~(\ref{Intro_Caputo}) reads $\mathcal{L}\{_0C_{t}^\alpha g(t)\}=\lambda^\alpha g(\lambda)-\lambda^{\alpha-1}g(t=0)$.
Having all this in mind, the inverse Laplace transform of Eq.~(\ref{2_ffpe_L}) then yields our main result:
\begin{eqnarray}
\label{2_ffpe}
_0C_{t_1t_2}^\alpha f(x_2,t_2;x_1,t_1)+{_0C_{t_2}^\alpha} f(x_2,t_2)\delta(x_1)+{_0C_{t_1}^\alpha} f(x_1,t_1)\delta(x_2)&=&L(x_2,x_1)f(x_2,t_2;x_1,t_1)\nonumber\\&&+h(0,t_2;0,t_1)\delta(x_2)\delta(x_1).
\end{eqnarray}
This is the two-point fractional diffusion equation, generalizing the single-time FDE to two-point probability distributions. Here, a generalization of the single-time Caputo operator to two times is introduced as:
\begin{eqnarray}
\label{2_caputo_fracdiff}
_0C_{t_1t_2}^\alpha g(t_2,t_1)&=&\frac{1}{\Gamma(1-\alpha)}[\Theta(t_2-t_1)t_1^{-\alpha}+\Theta(t_1-t_2)t_2^{-\alpha}]**\frac{\partial}{\partial t_2}\frac{\partial}{\partial t_1}g(t_2,t_1),
\end{eqnarray}
where the $**$ denote a double Laplace convolution with respect to $t_2$ and $t_1$. With the definition of the two-time fractional Caputo derivative Eq.~(\ref{2_caputo_fracdiff}) the general form of the diffusion equation is clearly visible. In addition, boundary terms occur which can be interpreted according to the underlying random walk. $_0C_{t_2}^\alpha f(x_2,t_2)\delta(x_1)$ and $_0C_{t_1}^\alpha f(x_1,t_1)\delta(x_2)$ describe the propagation when either $t_1=0$ or $t_2=0$. $h(0,t_2;0,t_1)\delta(x_2)\delta(x_1)$ is due to the non-zero probability that the random walker stays at the initial site. The analytical form of this term is obtained by performing the inverse Laplace transform of $h(0,\lambda_2;0,\lambda_1)$ (Eqs.~(\ref{2_hinitial_L})):
\begin{eqnarray}
h(0,t_2;0,t_1)&=&\Theta(t_2-t_1)h(0,t_2)+\Theta(t_1-t_2)h(0,t_1),
\end{eqnarray}
where $h(0,t_i)=t_i^{-\alpha}/\Gamma(1-\alpha)$. This expression agrees with the interpretation given above. For $t_2>t_1$, the probability of staying at the initial site until $t_2$ is given by the single-time distribution $h(0,t_2)$ and vice versa. As in the single-time case, the two-point FDE reduces to the form of its Markovian counterpart Eq.~(\ref{Joint_ODE}) for $\alpha\rightarrow 1$ (see the expression in Laplace space Eq.~(\ref{2_ffpe_L})). Also, the transformation Eq.~(\ref{path_integral}) reproduces this result, since for $\alpha\rightarrow 1$:  $h(s_2,t_2;s_1,t_1)=\delta(s_2-t_2)\delta(s_1-t_1)$ (see Eq.~(\ref{2_h_L})). The occurrence of additional boundary terms in Eq.~(\ref{2_ffpe}) can be considered as a general signature of non-Markovian processes and is even more prominent in the multiple-time case. A generalization to $n$-point probability distributions is obtained in a straightforward way along the lines outlined above.

\begin{figure}
\begin{center}
\includegraphics[width=16cm]{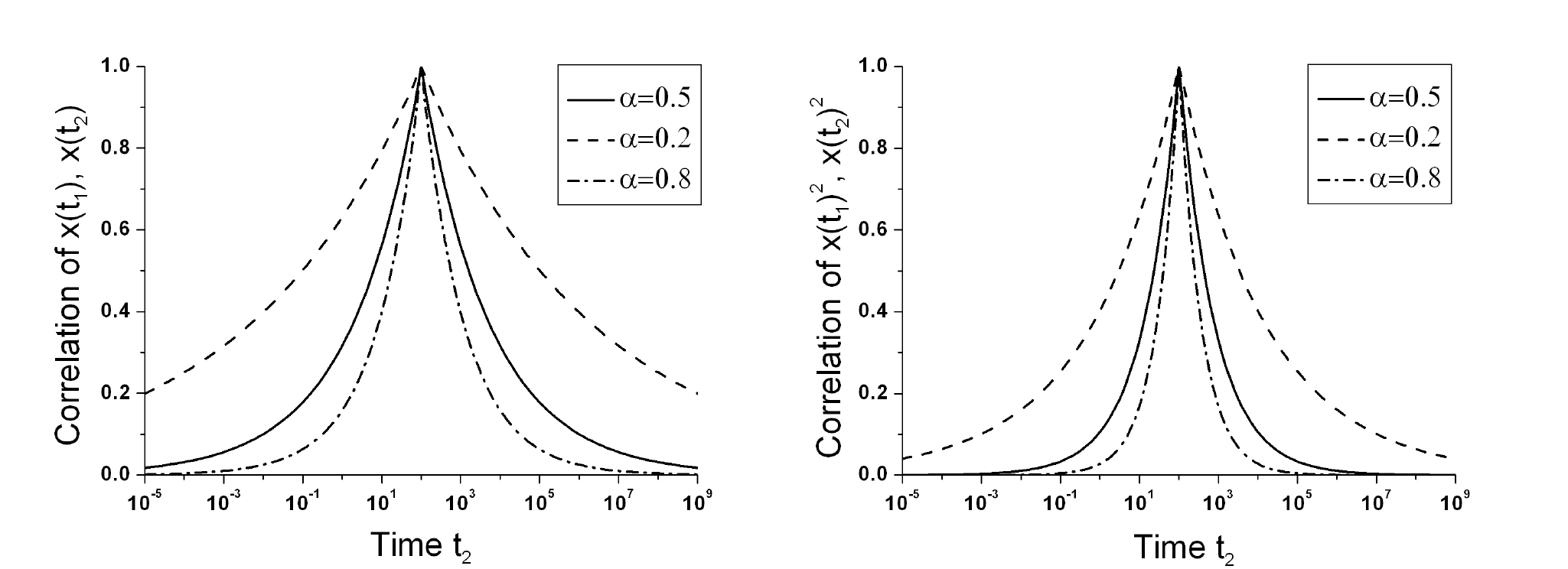}
\caption{\label{Correlation}Semi-logarithmic plot of the correlations of $x(t_1)$, $x(t_2)$ and $x(t_1)^2$, $x(t_2)^2$. Here, $t_1$ is fixed at $100$, setting the symmetry axis, and $\zeta=1$. All units are non-dimensional.}
\end{center}
\end{figure}

From Eq.~(\ref{2_ffpe}) the two-time moments of the subdiffusive CTRW can be calculated without invoking the properties of the inverse L\'evy-stable process $s(t)$. Consider e.g. the simplest moment $\left<x(t_2)x(t_1)\right>$. Multiplicating Eq.~(\ref{2_ffpe}) with $x_2$, $x_1$ and integrating from $-\infty$ to $\infty$ yields $_0C^\alpha_{t_2t_1}\left<x(t_2)x(t_1)\right>=\zeta$, 
where $\zeta$ is the strength of the stochastic force in the Langevin equation (\ref{Langevin1}). A solution of this equation can be calculated in Laplace space. Since we use the usual convention $X(t=0)=0$ the boundary terms due to the Laplace transformation of $_0C^\alpha_{t_2t_1}$ all vanish. One immediately obtains the solution $\mathcal{L}\{\left<x(t_2)x(t_1)\right>\}=\zeta/(\lambda_1\lambda_2(\lambda_1+\lambda_2)^\alpha)$, which reads in real space:
\begin{eqnarray}
\label{moment1}
\left<x(t_2)x(t_1)\right>&=& \Theta(t_2-t_1)\zeta\frac{t_1^\alpha}{\Gamma(\alpha+1)}+ \Theta(t_1-t_2)\zeta\frac{t_2^\alpha}{\Gamma(\alpha+1)}.
\end{eqnarray}
Due to the fact, that $f(x_2,t_2;x_1,t_1)$ is given as the transformation Eq.~(\ref{path_integral}) only even higher order moments are non-zero. They obey the general recursion formula (for non-negative integers $m$, $n$):
\begin{eqnarray}
_0C^\alpha_{t_2t_1}\left<x(t_2)^{2m}x(t_1)^{2n}\right>&=&m(2m-1)\zeta\left<x(t_2)^{2m-2}x(t_1)^{2n}\right>+n(2n-1)\zeta\left<x(t_2)^{2m}x(t_1)^{2n-2}\right>\nonumber\\
&&+4mn\zeta\left<x(t_2)^{2m-1}x(t_1)^{2n-1}\right>.
\end{eqnarray}
In the case $m=n=1$ we obtain:
\begin{eqnarray}
\label{moment2}
\left<x(t_2)^2x(t_1)^2\right>&=&\Theta(t_2-t_1)\zeta^2\left\{\frac{5t_1^{2\alpha}}{\Gamma(2\alpha+1)}+\frac{t_2^\alpha t_1^\alpha}{\Gamma(\alpha+1)^2}F\left(\alpha,-\alpha;\alpha+1;\frac{t_1}{t_2}\right)\right\}\nonumber\\
&&+\Theta(t_1-t_2)\zeta^2\left\{\frac{5\ t_2^{2\alpha}}{\Gamma(2\alpha+1)}+\frac{t_2^\alpha t_1^\alpha}{\Gamma(\alpha+1)^2}F\left(\alpha,-\alpha;\alpha+1;\frac{t_2}{t_1}\right)\right\}.
\end{eqnarray}
Here, $F(a,b;c;z)$ denotes the hypergeometric function (see e.g. \cite{Abram}). Figure \ref{Correlation} shows a semi-logarithmic plot of the correlations corresponding to Eq.~(\ref{moment1}) and Eq.~(\ref{moment2}) as functions of $t_2$ for three different $\alpha$-values. Both correlations exhibit a clear power law decay for $t_2\gg t_1$: $\propto t_2^{-\alpha/2}$ and $\propto t_2^{-\alpha}$ respectively. It would be interesting to determine these correlations from experiments.

We have demonstrated that a consistent generalization of the well-known single-time fractional diffusion equation to two-point pdfs can be derived on the basis of the coupled Langevin equations introduced by Fogedby as a representation of CTRWs. Special features of the two-point FDE are a two-time fractional differential operator of the Caputo-type and the occurrence of additional boundary terms. Its solution is expressed in terms of an integral transformation of the two-point pdf of the corresponding normal diffusion process. As in the single-time case the $\alpha\rightarrow 1$ limit reduces the FDE to its Markovian counterpart. Furthermore we derived recursion relations for arbitrary two-time moments of the subdiffusive CTRW. It should be noted that the occurrence of fractional derivatives is a consequence of the properties of the stochastic process $t(s)$ which determines the temporal behaviour of the CTRW. Thus the derivation of evolution equations, as presented in this paper, should apply to a whole class of systems, which can be described by two independent stochastic processes for $X$ and $t$. Here the simplest case has been solved, namely $X$ specified as a Wiener process. For this case, we have determined explicit expressions for the two-time moments, which could be readily compared with moments obtained from experiments.
An investigation of the multi-point statistics of other CTRW related processes, e.g. the anomalous diffusion of weakly damped inertial particles \cite{Friedrich2}, is left for future work.


\begin{thebibliography}{}
\bibitem{Montroll} E. W. Montroll and G. H. Weiss, \textit{J. Math. Phys.} \textbf{6}, 167 (1965).
\bibitem{Bouchaud} J.-P. Bouchaud and A. Georges, \textit{Phys. Rep.} \textbf{195}, 127 (1990).
\bibitem{Metzler1} R. Metzler and J. Klafter, \textit{Phys. Rep.} \textbf{339}, 1 (2000).
\bibitem{Shlesinger} M. F. Shlesinger, G. M. Zaslavsky, and J. Klafter, \textit{Nature} \textbf{363}, 31 (1993).
\bibitem{Metzler2} R. Metzler, E. Barkai, and J. Klafter, \textit{Phys. Rev. Lett.} \textbf{82}, 3563 (1999).
\bibitem{Podlubny} I. Podlubny, \textit{Fractional Differential Equations} (Academic Press, New York, 1999).
\bibitem{Schneider} W. R. Schneider and W. Wyss, \textit{J. Math. Phys.} \textbf{30}, 134 (1989).
\bibitem{Friedrich1} R. Friedrich, \textit{Phys. Rev. Lett.} \textbf{90}, 084501 (2003).
\bibitem{Metzler3} R. Metzler and J. Klafter, \textit{J. Phys. A} \textbf{37}, R161 (2004).
\bibitem{Barkai} E. Barkai, \textit{Phys. Rev. Lett.} \textbf{90}, 104101 (2003).
\bibitem{Bel} G. Bel and E. Barkai, \textit{Phys. Rev. Lett.} \textbf{94}, 240602 (2005).
\bibitem{Friedrich2} R. Friedrich, F. Jenko, A. Baule, and S. Eule, \textit{Phys. Rev. Lett.} \textbf{96}, 230601 (2006).
\bibitem{Barsegov} V. Barsegov and S. Mukamel, \textit{J. Phys. Chem. A} \textbf{108}, 15 (2004).
\bibitem{Baule} A. Baule and R. Friedrich, \textit{Phys. Rev. E} \textbf{71}, 026101 (2005).
\bibitem{Yang} H. Yang et al., \textit{Science} \textbf{302}, 262 (2003).
\bibitem{Min} W. Min et al., \textit{Phys. Rev. Lett.} \textbf{94}, 198302 (2005) .
\bibitem{Fogedby} H. C. Fogedby, \textit{Phys. Rev. E} \textbf{50}, 1657 (1994).
\bibitem{Abram} \textit{Handbook of Mathematical Functions}, edited by M. Abramowitz and C. A. Stegun (Dover, New York, 1972).


\end{thebibliography}
\end{document}